\documentclass[twocolumn,trackchanges]{aastex631}
\usepackage{graphicx}
\usepackage{float}
\usepackage{enumitem}
\usepackage{CJK}

\shortauthors{L. Wang et al.}

\defcitealias{li_formation_2016}{L16}

\graphicspath{{./}{figures/}}
\begin{document}
\begin{CJK*}{UTF8}{gbsn}

\title{The Origin of Young Stellar Populations in NGC 1783: Accretion of External Stars}

\correspondingauthor{Chengyuan Li}
\email{lichengy5@mail.sysu.edu.cn}

\author[0000-0003-3471-9489]{Li Wang (王莉)}
\affiliation{School of Physics and Astronomy, Sun Yat-sen University, Daxue Road, Zhuhai, 519082, People's Republic of China}
\affiliation{CSST Science Center for the Guangdong--Hong Kong--Macau Greater Bay Area, Zhuhai, 519082, People's Republic of China}

\author[0000-0001-9073-9914]{Licai Deng (邓李才)}
\affiliation{Key Laboratory for Optical Astronomy, National Astronomical Observatories, Chinese Academy of Sciences, Beijing 100101, People's Republic of China}
\affiliation{School of Astronomy and Space Science, University of Chinese Academy of Sciences, Beijing 100049, People's Republic of China}
\affiliation{Department of Astronomy, School of Physics and Astronomy, China West Normal University, Nanchong 637002, People's Republic of China}

\author[0000-0003-3389-2263]{Xiaoying Pang (庞晓莹)}
\affiliation{Department of Physics, Xi'an Jiaotong--Liverpool University, 111 Ren'ai Road, Dushu Lake Science and Education Innovation District, Suzhou 215123, Jiangsu Province, People's Republic of China}
\affiliation{Shanghai Key Laboratory for Astrophysics, Shanghai Normal University, 100 Guilin Road, Shanghai 200234, People's Republic of China}

\author[0000-0001-8713-0366]{Long Wang (王龙)}
\affiliation{School of Physics and Astronomy, Sun Yat-sen University, Daxue Road, Zhuhai, 519082, People's Republic of China}
\affiliation{CSST Science Center for the Guangdong--Hong Kong--Macau Greater Bay Area, Zhuhai, 519082, People's Republic of China}

\author[0000-0002-7203-5996]{Richard de Grijs}
\affiliation{School of Mathematical and Physical Sciences, Macquarie University, Balaclava Road, Sydney, NSW 2109, Australia}
\affiliation{Astrophysics and Space Technologies Research Centre, Macquarie University, Balaclava Road, Sydney, NSW 2109, Australia}
\affiliation{International Space Science Institute--Beijing, 1 Nanertiao, Zhongguancun, Hai Dian District, Beijing 100190, People's Republic of China}

 \author[0000-0001-7506-930X]{Antonino P. Milone}
 \affil{Dipartimento di Fisica e Astronomia ``Galileo Galilei'', Univ. di Padova, Vicolo dell'Osservatorio 3, Padova, IT-35122, Italy}
 \affil{Istituto Nazionale di Astrofisica - Osservatorio Astronomico di Padova, Vicolo dell'Osservatorio 5, Padova, IT-35122, Italy}
 
\author[0000-0002-3084-5157]{Chengyuan Li (李程远)}
\affiliation{School of Physics and Astronomy, Sun Yat-sen University, Daxue Road, Zhuhai, 519082, People's Republic of China}
\affiliation{CSST Science Center for the Guangdong--Hong Kong--Macau Greater Bay Area, Zhuhai, 519082, People's Republic of China}

\begin{abstract}
The presence of young stellar populations in the Large Magellanic Cloud cluster NGC 1783 has caught significant attention, with suggestions ranging from it being a genuine secondary stellar generation to a population of blue straggler stars or simply contamination from background stars. Thanks to multi-epoch observations with the {\sl Hubble Space Telescope}, proper motions for stars within the field of NGC 1783 have been derived, thus allowing accurate cluster membership determination. Here, we report that the younger stars within NGC 1783 indeed belong to the cluster, and their spatial distribution is more extended compared to the bulk of the older stellar population, consistent with previous studies. Through $N$-body simulations, we demonstrate that the observed characteristics of the younger stars cannot be explained solely by blue straggler stars in the context of the isolated dynamical evolution of NGC 1783. Instead, accretion of the external, low-mass stellar system can better account for both the inverse spatial concentration and the radial velocity isotropy of the younger stars. We propose that NGC 1783 may have accreted external stars from low-mass stellar systems, resulting in a mixture of external younger stars and blue straggler stars from the older bulk population, thereby accounting for the characteristics of the younger sequence.
\end{abstract}

\keywords{Star clusters (1567) --- Stellar kinematics (1608) --- Blue straggler stars (168) --- $N$-body simulations (1083)}

\section{Introduction} \label{sec:intro}

Star clusters were traditionally thought to originate from a single episode of star formation, resulting in simple stellar populations (SSPs) that shared the same age and chemical composition \citep{bruzual2010star}. Nevertheless, accumulating evidence from spectroscopic and photometric studies indicates that globular clusters (GCs) harbor multiple stellar populations (MPs) with distinct chemical compositions. These populations exhibit significant differences in
specific spectral lines or passbands and demonstrate detectable
multiple patterns in their color--magnitude diagrams (CMDs), such as multiple (or broadened) main sequences (MSs) and/or red-giant branches (RGBs), or elongated horizontal branches (HBs) \citep[e.g.,][]{lee1999, Piotto_2007, Milone_2008, Carretta2009, Bellini_2013, Li_2014, 10.1093/mnras/stw611, Pancino2017, Lagioia_2021}.

Various scenarios have been proposed to explain the presence of
MPs. Most proposals suggest that second-generation (SG) stars (which thus are chemically enriched) should form in the central regions of more diffuse first-generation (FG) systems
\citep{10.1111/j.1365-2966.2008.13915.x}. \citet{Dalessandro_2019} discovered that clusters with young dynamical ages preferentially contain centrally concentrated SG populations, while clusters with older dynamical ages exhibit both populations being fully mixed. However, existing theories cannot explain dynamically young GCs with diffuse SG populations \citep{Dalessandro_2019, leitinger_wide-field_2023}. 
Stellar kinematics, including rotation and velocity dispersion, also offers insights into MP
formation. \citet{10.1093/mnras/stw2812} found differential rotation among MPs in M13. Those chemically anomalous stars were found to exhibit more radially anisotropic velocity distributions than normal stars \citep{Richer_2013, Bellini_2015, Libralato_2018, Cordoni_2020}. 
However, because of the generally old ages of GCs ($\ge$10 Gyr), whether there is an age difference between such proposed MPs in these scenarios cannot be detected with current techniques. 

If MPs truly represent different stellar generations, younger clusters should show multiple generations (MGs) of stars with detectable differences in age. \citet[][hereafter \citetalias{li_formation_2016}]{li_formation_2016} found that three massive, intermediate-age Magellanic Cloud (MC) clusters---NGC 1783, NGC 1806, and NGC 411---contain younger generations of stars populating distinct sequences within their CMDs. Those younger-population stars are less centrally concentrated than the bulk population stars in these clusters, which is at odds with the results from previous theoretical studies of MP formation \citep[e.g.,][]{10.1111/j.1365-2966.2008.13915.x}. These findings suggest that the origin of the MPs in these clusters could be different from that in old GCs. \citetalias[]{li_formation_2016} suggested that this could imply that star clusters are capable of accumulating gas from their environments, resulting in new star formation, although this assertion remains speculative. \citet{hong_dynamical_2017} advocated the association of younger sequences with host star clusters. They suggested that such younger sequences probably originate from minor mergers of clusters. However, a major challenge is that the lack of proper motion information for stars in the cluster region at that time made it difficult to determine to what extent these stars were actually members of the star clusters of interest. Therefore, this result was challenged by \citet{cabrera-ziri_no_2016}, who suggested that the observed younger
populations are, in fact, field stars that were not adequately
subtracted.

With the increasing accumulation of observational data from the {\sl Hubble Space Telescope} ({\sl HST}), it has become possible to differentiate member stars from field stars within MC clusters based on their proper motions. Employing this technique, \citet{milone_hubble_2023} was the first to identify high-confidence member stars for 13 MC clusters, including the Large Magellanic Cloud (LMC) cluster NGC 1783. Their results support that the younger stars detected by \citetalias[]{li_formation_2016} are indeed cluster members. However, at that time, because of the lack of direct comparisons to $N$-body simulations, the origin of these young population stars is still unclear, although they suggest that these young stars are possibly blue straggler stars (BSSs). 

In this work, we aim to determine the origin of the young stellar populations detected in NGC 1783 by comparing their observational characteristics with $N$-body simulations. Using the unsupervised machine learning method based on neural network algorithms, we revisit the findings of \citet{milone_hubble_2023} and confirm their conclusion that the young stellar population identified by \citet{li_formation_2016} are indeed genuine cluster members. Section \ref{sec: model} introduces the method and data reduction. In Section \ref{sec: results}, we present our main results, including the relative radial distributions of the younger and bulk population stars in NGC 1783, as well as their kinematic patterns, and we compare them to $N$-body simulations. We discuss these results and some limitations in that same section. Finally, we provide discussions and a summary in Section \ref{sec: conclusions}.

\section{Observations and $N$-body Simulations} \label{sec: model}
 \begin{figure*}[ht!]
 \epsscale{1.15}
 \plotone{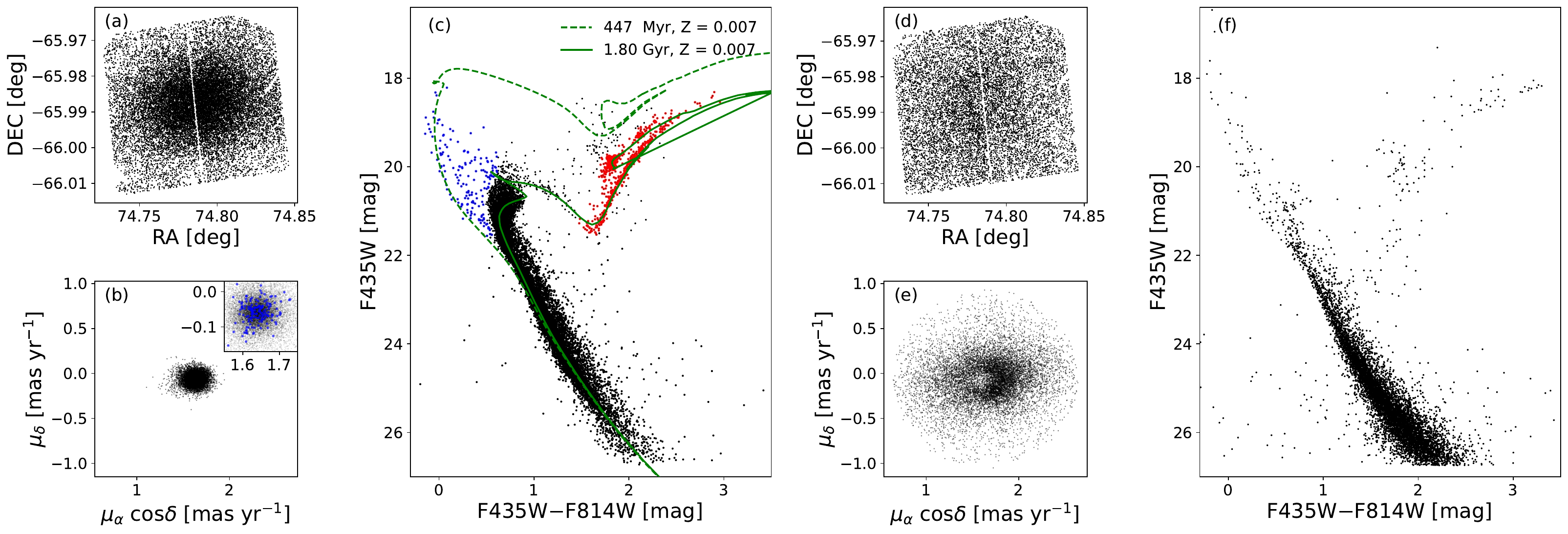}
 \caption{ (a) Spatial distribution and (b) proper motion diagram of member stars in the field of NGC 1783. 
 (c) Field-star decontaminated CMD of NGC 1783. Blue and red dots represent young-sequence stars and the corresponding reference stars used for comparison, respectively. The green solid and dashed lines are best-fitting PARSEC isochrones with different ages labeled \citep{10.1111/j.1365-2966.2012.21948.x}. 
The inset in panel (b) highlights the proper motion distribution of young-sequence stars. 
(d-f) As panels (a-c), but for field stars, which are then subtracted from the raw stellar catalog.}
 \label{f1}
 \end{figure*}

We directly use the NGC 1783 stellar catalog of \citet{milone_hubble_2023}, which includes both astrometry and photometry for each star, derived from high-precision multi-epoch {\sl HST} images. Following \citetalias[]{li_formation_2016}, we use photometry in the F435W and F814W bands of the Ultraviolet and Visual
Channel of the Wide Field Camera 3 (UVIS/WFC3). The methods applied to estimate proper motions and the dataset details are outlined in \citet{milone_hubble_2023}, and observational information is presented in their Table 4.

We use the unsupervised machine learning method, {\tt\string StarGO}\footnote{\url{https://github.com/zyuan-astro/StarGO-OC}}, to identify the NGC 1783 member stars. The algorithm is founded on the self-organizing map, which enables the projection of high-dimensional data onto a two-dimensional (2D) neural network while preserving the intrinsic topological structures of the dataset. Consequently, stars clustered in the high-dimensional space are associated with neurons
grouped on the 2D map. We apply {\tt\string StarGO} to map a four-dimensional (4D) dataset (R.A. $\alpha$, Dec. $\delta$, $\mu_{\alpha} \rm{cos}\delta$, $\mu_{\delta}$) onto a 2D neural network. In brief,
we initially constructed a 2D network to match the number of neurons with the number of stars in our input sample. Each neuron had a weight vector of equivalent dimensionality to the input vector and underwent
an adjustment to approximate the input vector associated with a specific star as stars were fed, one at a time, to all neurons. A single iteration was considered complete when all stars had been
processed by the neurons once. The entire learning cycle was repeated for 400 iterations until convergence of the weight vectors was achieved. 

The cluster membership is determined as follows. Neurons exhibiting similarity in their 4D weight vectors were clustered on the 2D neural network and represented stars sharing similar spatial and kinematic attributes (member stars). The magnitude of the disparity in weight vectors between neighboring neurons is denoted as $u$. Member stars in NGC 1783 correspond to a local minimum among the $u$ values, associated with neurons with a $u$ value below a specific threshold. The best-fitting bi-Gaussian function can be used to estimate the membership probabilities to fit the bimodal log $u$ histogram distribution. Notably, a component with smaller $u$ values represents potential member stars, while another with larger $u$ values designates field stars. Although this is the first time we apply {\tt\string StarGO} to determine MC cluster member stars, we emphasize that {\tt\string StarGO} has already achieved success in identifying members of both open clusters \citep{Tang_2019, Pang_2020, Pang_2021a, Pang_2021b, Pang_2022} and stellar streams \citep{Yuan_2020a, Yuan_2020b}.

In this work, where the projected stellar number density reaches its maximum is defined as the cluster center, and the so derived coordinates are $\alpha_{\rm J2000}$ = $04^{\rm h}59^{\rm m}08^{\rm s}{.}76$ and $\delta_{\rm J2000}$ = $-65^{\circ}59^{\prime}15^{\prime\prime}{.}38$. 
The empirical King model \citep{1962AJ.....67..471K} and a constant number density of the background field population are adopted to fit the observed stellar number-density profile in the field of view of NGC 1783: 
\begin{equation}
\rho(r)=k\left[\frac{1}{\sqrt{1+\left(r / r_{\mathrm{c}}\right)^2}}-\frac{1}{\sqrt{1+\left(r_{\mathrm{t}} / r_{\mathrm{c}}\right)^2}}\right]^2+b.
\label{eq1}
\end{equation}
Where $r_{\rm c}$ and $r_{\rm t}$ are the core and tidal radii, respectively, $b$ is the background field number density, $k$ is a normalization coefficient, $\rho$ is the number density, and $r$ is the distance from a star to the cluster's center. 
Bright stars with F435W $\leq$ 22 mag were taken into consideration, whose photometric completeness is better than 90\%. 
The threshold for $u$ is controlled by subtracting the expected number of field stars of F435W brighter than 22 mag as estimated by a best-fitting King model from the raw catalog.

The error in proper motions remains to be the primary uncertainty for membership. We performed 100 iterations of {\tt\string StarGO}, with each run incorporating randomly sampled proper motion errors following Gaussian distributions. We confirm that the detected younger stellar population in
\citetalias[]{li_formation_2016} are genuine cluster members of NGC 1783
according to their high cluster membership probabilities (mostly $\geq 95\%$), as shown in Figure \ref{f1}.
Finally, $152 \pm 4$ younger stars are selected as candidate members of NGC 1783.

Accurate cluster membership is crucial for further unbiased analysis of NGC 1783. We estimated the dynamical masses using Equation (1) in \citet{2006MNRAS.369.1392F} by deriving proper motion velocity dispersions of member and field stars in Figure \ref{f1}(b, e). The obtained masses are roughly comparable to that of NGC 1783 and the LMC field. 
It is not surprising to see such substructures in the spatial distribution of the field stars on the left-hand side of Figure \ref {f1}(d) because it would be challenging to disentangle field stars from cluster stars fainter than F435W = 23.5mag. 
This is not crucial since we only focus on stars brighter than F435W = 22mag in this work.
We also applied our cluster member selection method to numerically simulated clusters with homogeneous fields to verify the reliability of the member stars. We find that we can correctly derive $\sim$ 90\% cluster genuine members, including at least $\sim$ 95\% bright members, using our method, demonstrating the robustness of our cluster membership identification.

To constrain the origin of younger-population stars in NGC 1783, we
run a realistic numerical simulation using the high-performance
$N$-body code {\tt\string PETAR}\footnote{\url{https://github.com/lwang-astro/PeTar}}. {\tt\string
  PETAR} allows us to efficiently mimic the evolution of a massive
stellar system containing up to $10^5$ particles with a large fraction
of binaries, up to unity
\citep{10.1093/mnras/staa1915,wang_impact_2022}. To accurately follow
the dynamical and stellar evolution of both single stars and binary
systems, the recently updated single and binary stellar evolution
codes, {\tt\string SSE} and {\tt\string BSE}
\citep{10.1093/mnras/291.4.732,10.1046/j.1365-8711.2000.03426.x,10.1046/j.1365-8711.2002.05038.x,refId1},
were incorporated in {\tt\string PETAR} to simulate wind mass loss,
stellar type changes, mass transfer, and binary mergers. Although suffering from the uncertain parametric treatment of the complicated physical processes, the massive BSSs still obey the trend of the most massive particles during the dynamical evolution of the host cluster.

We use the newly updated version of the star cluster initial model
generator code {\tt\string
  MCLUSTER}\footnote{\url{https://github.com/lwang-astro/mcluster}}
\citep{10.1111/j.1365-2966.2011.19412.x,10.1093/mnras/sty2232} to
generate an isolated cluster. The initial total mass is $2\times10^5$
$M_{\odot}$. The initial half-mass radius is 1.4 pc, including all
stellar components in three-dimensional (3D) space, estimated empirically
from the present-day observed data (the current total mass and the
half-mass radius of NGC 1783 are $1.78\times 10^5$ $M_{\odot}$ and
9.0 pc; \citealt{goudfrooij_population_2011}). The cluster
metallicity is $Z =0.01$ ($Z_{\odot}=0.02$). The initial particle
masses were randomly sampled from a Kroupa-like initial mass function
\citep[IMF;][]{10.1046/j.1365-8711.2001.04022.x} covering a mass range
of 0.08--150 $M_{\odot}$. Stars' 3D positions and velocities were
randomly sampled from a Plummer density profile
\citep{1974A&A....37..183A}. Previous studies suggested that a
primordial binary fraction up to 100\% better restores the observed
binary fractions inside and outside half-mass radii of GCs
\citep{Leigh2015TheSO} and in observed GC CMDs
\citep{belloni_initial_2017}. Our adopted simulation model is thus
initialized by adopting a 100\% primordial binary fraction. Our
simulation does not consider primordial mass segregation or tidal fields. We will later explain why we ignore the effect of an external tidal field in this simulation.

\begin{figure}[ht!]
\epsscale{1.2}
 \plotone{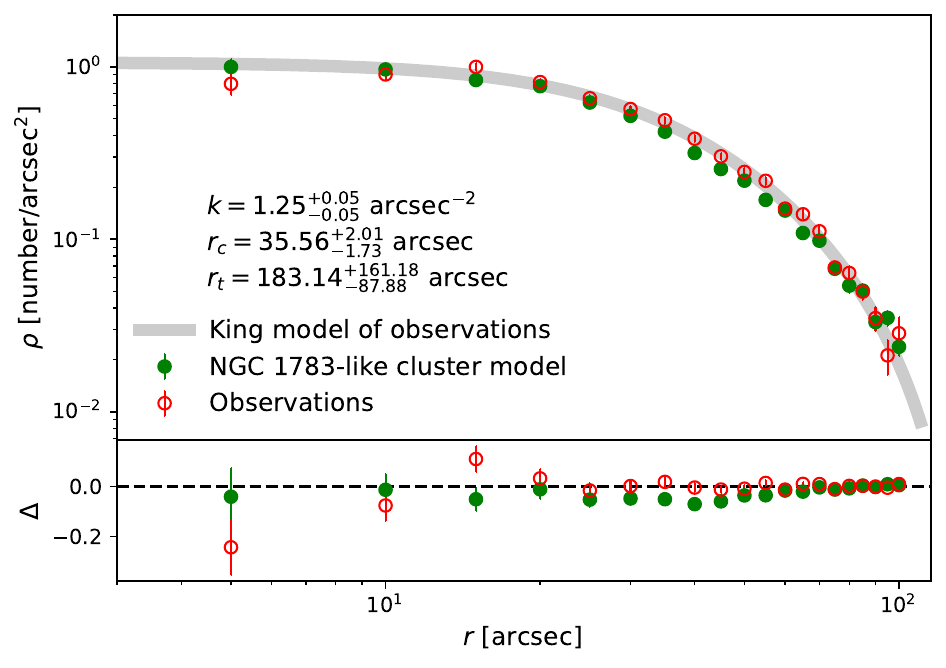}
 \caption{The normalized radial number density profiles for NGC 1783 observations  (background-subtracted, red circles) and best-fitting $N$-body model (NGC 1783-like cluster model, green dots). The gray curve is the best-fitting King model of the observations of NGC 1783. The bottom panel shows the residuals between the best-fitting King model and density profiles.}
 \label{f2}
 \end{figure}

In summary, we used the {\tt\string PETAR} code to mimic an NGC
1783-like cluster, including full and realistic stellar dynamics
and evolution details. 
This implies that our simulated cluster possesses a density profile (Figure \ref{f2}) and age that closely approximate the real
  observation. The remaining total mass is $\sim 1.4\times10^5$
  $M_{\odot}$. We also simulated the merger of two star clusters using the {\tt\string PETAR} code. However, for the sake of
time efficiency, we conducted this merger simulation in a qualitative
manner. The specific parameter settings will be directly addressed in
Section \ref{subsec: merger}.


\section{Main Results} \label{sec: results}

 \begin{figure*}[ht!]
 \plotone{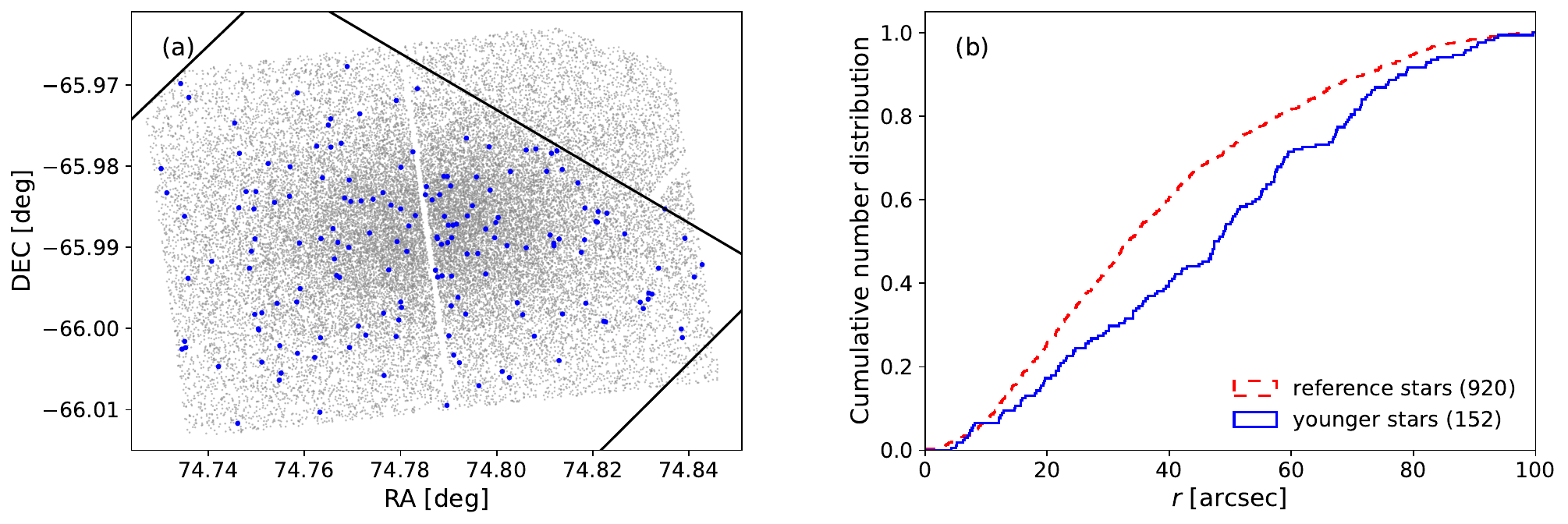}
 \caption{(a) Spatial distribution of young-sequence stars (blue dots). The black line represents the field for the F435W observation.
 (b) Cumulative radial distributions of young-sequence stars (blue solid line) and reference stars (red dashed line) in NGC 1783. The number of younger and reference stars selected in NGC 1783 are also indicated.}
 \label{f3}
 \end{figure*}

  \begin{figure*}[ht!]
 \plotone{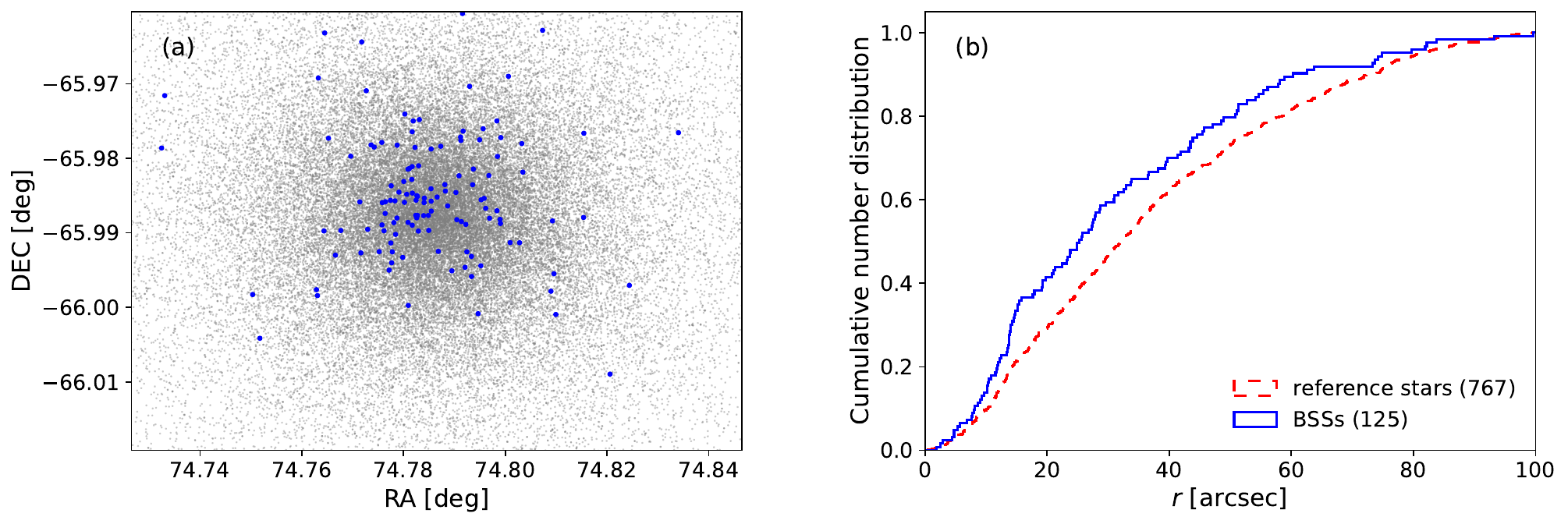}
 \caption{As Figure \ref{f3}, but for stars in the NGC 1783-like cluster model. In panel (a), the blue dots are all BSSs, and the blue solid line in panel (b) indicates their cumulative radial distribution.}
 \label{f4}
 \end{figure*}
 
\subsection{Blue Straggler Stars}

Analogously to \citetalias[]{li_formation_2016}, we first reanalyze
the cumulative radial distributions of the young population stars
compared with those of our reference stars with similar luminosities
(mostly RGB and red-clump (RC) stars). We only selected these evolved
stars as the reference population because those stars represent the
dominant older population and have the highest observational
completeness. We found that the stars making up the younger population
are significantly less centrally concentrated than the dominant older
population of cluster members (see Figure \ref{f3}), consistent with
previous result published by \citetalias[]{li_formation_2016} and
\citet{2024Mohandasan}.

Using the $N$-body simulation, we aim to test if this simulated
cluster itself can produce a younger sequence through binary
interactions (i.e., through the production of BSSs) with the observed
spatial distribution. Similarly, we compare the radial profiles of
the BSSs and reference stars in this NGC 1783-like cluster model, as
illustrated in Figure \ref{f4}. The trend that BSSs are marginally
more centrally concentrated than the normal evolved stars contradicts observations of younger stars.

We can also estimate a mass-segregation timescale \citep[defined as
  in][]{1987degc} for BSSs of a given mean mass ($\sim$2 $M_{\odot}$)
in the NGC 1783-like cluster model. The resulting mass-segregation
timescale is $\sim$479 Myr, significantly younger than the current age
of the model. Thus, the BSSs in our model are expected to have had
enough time to mass segregate dynamically, as is readily apparent in
Figure \ref{f4}(b). In addition, our simulation incorporates the
influence of black holes (BHs), which can decrease or completely
suppress mass segregation \citep{baumgardt_global_2017}, possibly implying that
the degree of mass segregation in the NGC 1783-like cluster model is a
lower limit. Regardless of the evolutionary timescale, among our model
results, there is no snapshot where BSSs have an inverse radial
distribution with a similar degree of mass segregation as in our
canonical simulation.

Therefore, NGC 1783 cannot generate the observed spatial
  distribution of the younger sequence through internal dynamics,
  although our model evolving from a 100\% primordial binary fraction
  has produced a similar number of BSSs (125) as younger stars
  (152). BSSs cannot be entirely responsible for the presence of these
  younger stars. Indeed, we cannot entirely dismiss the possibility
  that NGC 1783 possesses special initial conditions, thereby leading
  to the observed inverse radial spatial distribution of
  BSSs. However, our current understanding of the dynamical evolution of clusters makes it difficult to conceive what initial conditions could give rise to a distribution of BSSs that matches our observations. A reasonable explanation is that the extended nature of the younger population may imply an external origin.

\subsection{Cluster Merger \label{subsec: merger}}

\begin{figure*}[ht!]
   \begin{minipage}{\textwidth}  \centerline{\includegraphics[width=\textwidth]{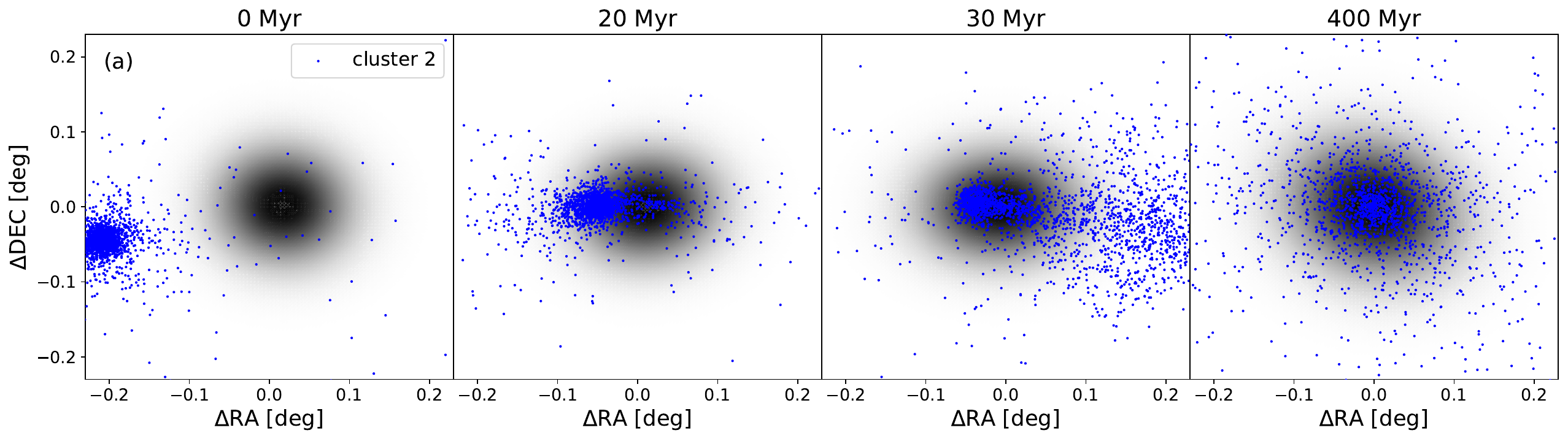}}
 \centerline{\includegraphics[width=\textwidth]{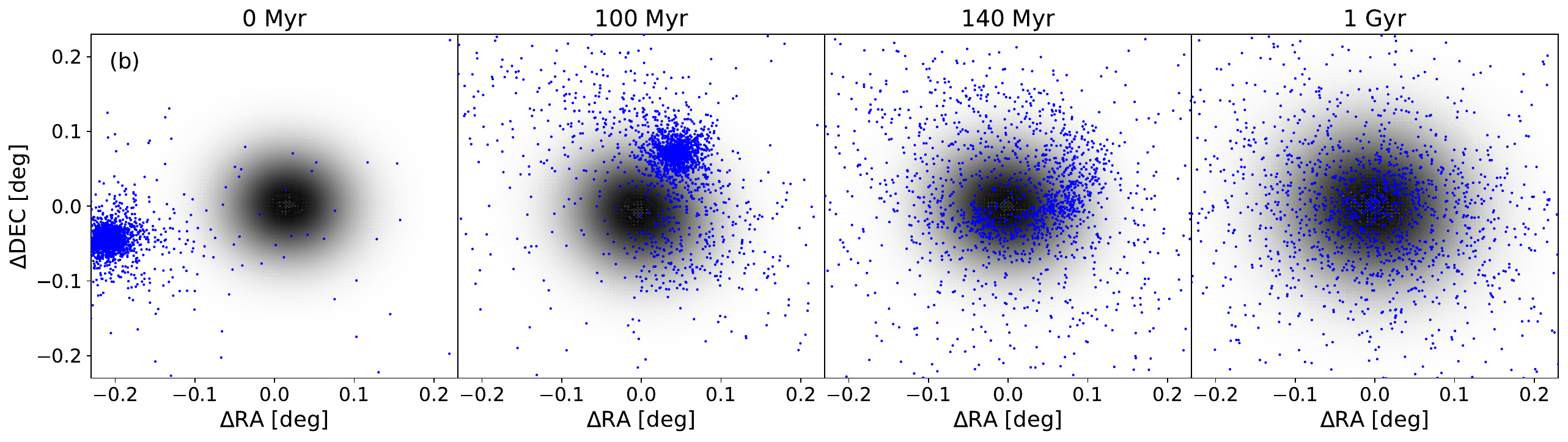}}
	\end{minipage}
	\caption{(a) In the head-on merger model: Relative positions of visible stars in cluster 1 (gray shades) and cluster 2 (blue dots) at different evolutionary timescales with respect to the cluster center. (b) As panel (a), but for the spiral merger model. Cluster 2 has been accreted by cluster 1 along an inward spiral path.}
 \label{f5}
\end{figure*}

A merger scenario has been proposed to explain the inverse radial
distributions between 1P and 2P stars in GCs by many observational and theoretical studies
\citep[e.g.,][]{2010ApJ...722L...1C, hong_dynamical_2017}. Mergers between two clusters occur only when their relative velocities are
smaller than (or of the same order of magnitude as) their velocity dispersions. 
Two clusters will be more likely to have a sufficiently low relative velocity to merge if they formed in a small dwarf galaxy like the MCs or the same molecular cloud \citep{10.1093/mnras/stw1397}.
Therefore, we consider that the younger sequence in NGC 1783 may originate from accretion of the other small, younger stellar system.
\citet{hong_dynamical_2017} confirmed, using numerical simulations,
that the minor merger scenario reproduces the observed inverse radial trends
of younger generations in NGC 411 and NGC 1806. Moreover, their younger
stellar population from the merger scenario exhibited noticeable
velocity anisotropy and rotational features.

We simulated two minor merger events and investigated the resulting
observational characteristics, including the radial distributions and
kinematic properties. We simulated two clusters, marked as cluster 1 and cluster
2, with masses of $3\times10^4$ $M_{\odot}$ and $6\times10^3$
$M_{\odot}$, respectively. 
The model clusters contain 50,000 and
10,000 stars, respectively, and their half-mass radii are 7 pc and 3 pc, adopting the cluster mass -- half-mass radius at birth time from \citet{marks_inverse_2012}.
{To avoid unbearable computational costs while keeping the goal of reproducing the main observations, we used such parameters and, at the same time, ignored the tidal fields and primordial binary stars during the simulation\footnote{Executing this simulation on our server, equipped with two Intel Xeon processors with 40 cores and 80 threads, totally takes approximately five months.}.}

\begin{figure}[ht!]
\epsscale{1.2}
 \plotone{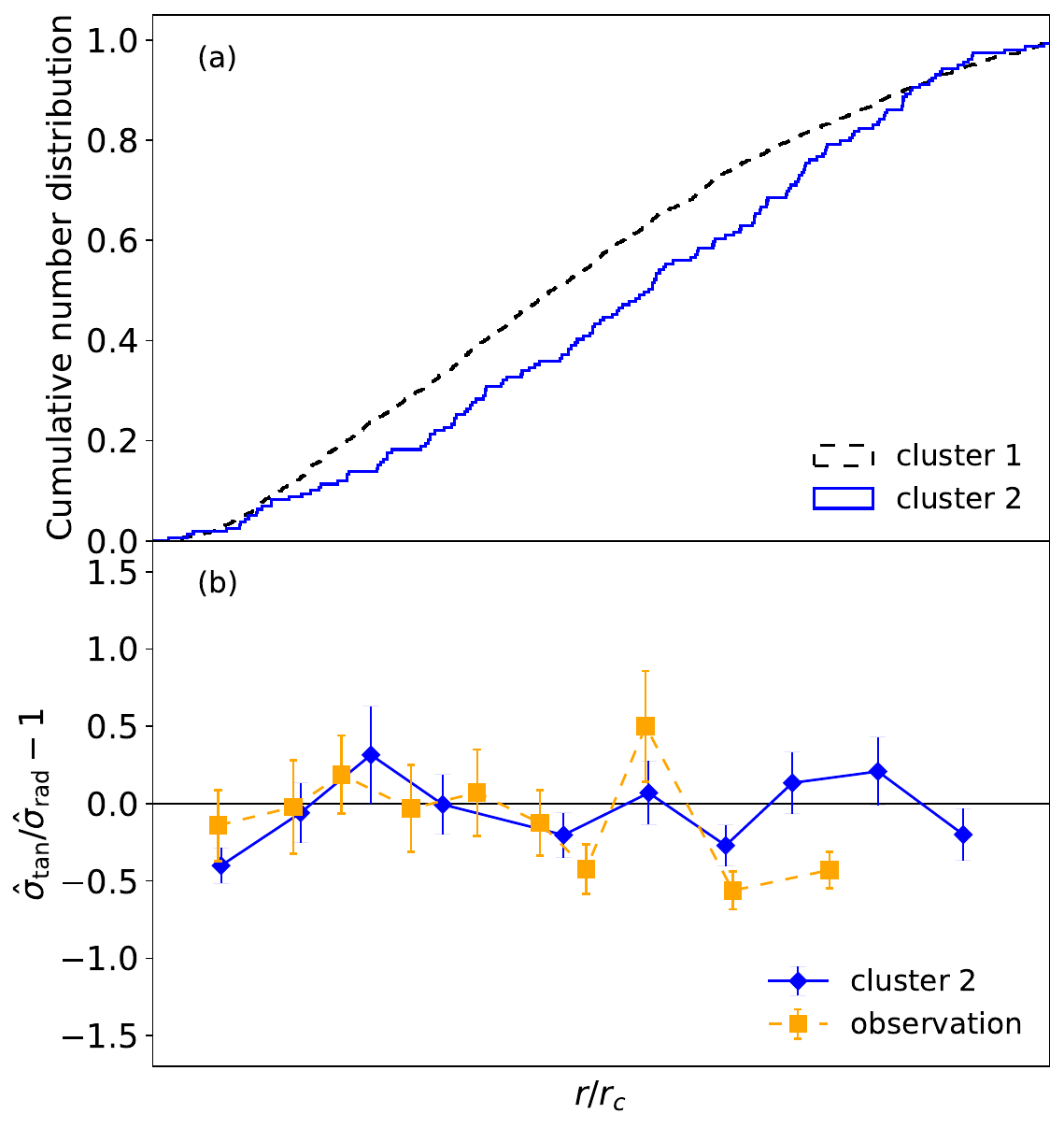}
 \caption{In the spiral merger model at the evolutionary timescale of 1.0 Gyr: (a) Cumulative radial distributions of visible bright stars in cluster 1 (black dashed line) and cluster 2 (blue solid line); 
(b) Deviation from tangential-to-radial isotropy ($\hat{\sigma}_{\rm tan}/\hat{\sigma}_{\rm rad}-1$) of visible bright stars in cluster 2 (blue diamonds) and younger stars in NGC 1783 (orange squares). \label{f6}}
 \end{figure}

The initial relative position of cluster 1 with respect to cluster 2 is ($-$20, $-$20, 20) pc. The subsequent evolutionary time starts from such initial geometry. Consequently, quite some time is required for the two clusters to approach each other and begin the actual merging process.
We conducted two particular simulations of the merger scenario by varying the initial relative velocities between cluster 1 and cluster 2: (1) A head-on merger: setting the initial relative velocity to be zero. In this case, cluster 2 will rapidly fall into the center of cluster 1, as in Figure \ref{f5}(a). Therefore, most stars from cluster 2 are less likely to spread throughout the outer region of the original cluster. (2) A spiral merger scenario: setting the initial velocity vector of cluster 2 relative to cluster 1 to be (1, 1, 1) km ${\rm s^{-1}}$, with the $v_z$ component oriented perpendicularly to the line connecting their centers. Cluster 2 merges with cluster 1 along a spiral path, as shown in Figure \ref{f5}(b). In this case, the stars of cluster 2 predominantly orbit the periphery of cluster 1, leading to a significant increase in the core radius of the final merged cluster compared with that of the stand-alone cluster 1 (evolving independently). The accreted stars exhibit a prolonged period of less segregation in their spatial distribution. Particularly during the early stages of the merger, this inverse spatial distribution is considerably more significant than observed for NGC 1783, leading to a misinterpretation as a distribution of field stars. In Figure \ref{f6}(a), the spiral merger model at the evolutionary timescale of 1.0 Gyr reproduces a similar inverse radial distribution found in the observations. Complete spatial mixing of both clusters will be reached once the whole cluster has undergone significant dynamical evolution. NGC 1783 is also one of the clusters with the largest core radii ($r_{\rm c}$ = 35.56 arcsec $\simeq$ 8.53 pc in our study) compared to its counterparts with similar total masses and ages \citep{rubele_star_2013, 2024Mohandasan}. The inverse radial distribution of younger stars and the large core radius suggest the possibility of a past merger event occurred in NGC 1783. 

We then explore the possibility of using the observed kinematic data to constrain the merger event further. We define $\hat{\sigma}_{\rm tan}$ and $\hat{\sigma}_{\rm rad}$ as the proper motion dispersions in the radial and tangential directions with respect to the cluster center, respectively. If the stellar motion is isotropic, the dispersions in these two orthogonal directions should be nearly equal, with the deviation from isotropy ($\hat{\sigma}_{\rm
  tan}/\hat{\sigma}_{\rm rad}-1$) approaching zero.
As \citet{hong_dynamical_2017} found, younger, accreted stars would exhibit a notable velocity anisotropy in a minor merger event. In our spiral merger model, we investigated the deviation from isotropy as a function of radial distance. During the short initial stage of the merger event, the accreted stars exhibit a pronounced radial anisotropy in the outer periphery of cluster 1. This radial velocity anisotropy is long-lasting and gradually diminishes. At $\sim$1.0 Gyr, the dissipation of the radial velocity anisotropy and the existence of an inverse spatial distribution of accreted stars are consistent with our observational findings, as shown in Figure \ref{f6}.

\subsection{The limitations of models}\label{sec: limitations}

The current simulation of the merger of two star clusters is rather qualitative. The mass ratio ($M_{\rm cluster\, 2}/M_{\rm cluster\, 1} \sim$20\%) employed in our merger simulation is relatively high. We note that if reducing the initial mass of cluster 2 to achieve a mass ratio of $\sim$10\%, the timescale ($\sim$1.0 Gyr) exhibiting similar observational features with NGC 1783 discussed in Section \ref{subsec: merger} decreases to $\sim$600 Myr. As such, one should be cautious when extrapolating our results to NGC 1783-like clusters. If cluster 1 were assumed to mimic NGC 1783, one would expect that the merger should eventually exhibit the current observational characteristics; however, the timescale involved would be different then. 

Here, we take a brief estimation of this timescale. The dynamical evolution of the merger system from initiation to the final stable spheroidal symmetry depends on violent relaxation and phase mixing processes, the timescales of which are generally dependent on the crossing time given by
\begin{equation}
t_{\rm cross} = \frac{2R}{v} = \frac{2R}{\sqrt{GM/R}} \propto R^{3/2}M^{-1/2}.
\end{equation}
Where $R$ represents the size of the star cluster, typically the tidal radius. 
Increasing the mass of cluster 1 to the mass of NGC 1783-like (and consequently reducing the mass ratio between cluster 1 and cluster 2) would accelerate the merger process. However, as the mass of cluster 1 increases, if the relative initial velocities of cluster 2 remain constant, it tends to fall towards the central region of cluster 1, thereby failing to reproduce the observed spatial distribution of younger population stars. Therefore, the initial velocity of cluster 2 needs to be higher accordingly, which will, in turn, raise the merging timescale. We expect that increasing the mass of Cluster 1 and raising the initial velocity of Cluster 2 relative to Cluster 1 will lead to a change of timescale in the range of several hundred million years. More realistic simulation requires a significant increase in computing power, and this will be addressed in future works.

For clarity, this study has assumed no tidal field from the Milky Way Galaxy and the LMC. NGC 1783 is situated in the LMC, and LMC is approximately 50 kpc away from the Galactic Center. First of all, the impact of the tidal field from the Milky Way on the cluster can be neglected at this distance. Moreover, the impact of the LMC tidal field on NGC 1783 is not known, although the projected distance between the cluster and the LMC center can be measured. To estimate the impact of the LMC field on NGC 1783, we assume that LMC is a point mass and the cluster is situated at a fixed distance. We calculate the ratio between the tidal force of LMC acting on the cluster and the gravitational force exerted by the LMC, obtaining a maximum value of $\sim$0.02. Therefore, the LMC's gravitational potential plays a minor role in the merger process, but may have subtle effects on the secular dynamical evolution of the merged system.

In general, low-mass stars in the outer regions of the cluster tend to acquire larger velocities and are more likely to be stripped away from the cluster. For the spatial distribution of BSSs in the central region (within $\sim0.5 r_{\rm t}$) of the simulated NGC 1783-like cluster model, the influence of the LMC tidal field appears to be limited. Therefore, we expect that the inclusion of an external tidal field would enhance the mass segregation effect of BSSs, thereby providing additional evidence against the hypothesis that all observed young population stars are composed exclusively of BSSs.

Considering the impact of LMC's tidal fields on the secular evolution of the system, more complicated simulations involving fine-tuning parameters describing the merger will be needed, which is outside the scope of the current paper. With the condition that the tidal field due to LMC is minor and is therefore neglected, the young stellar populations in NGC 1783 can be produced through external mergers.

\section{Summary and Discussion} \label{sec: conclusions}

In this work, we report that NGC 1783 harbors genuine younger population stars and discuss their origin by comparing observational characteristics with $N$-body simulations. The main results of the current study clearly indicate that one should be cautious when categorizing stars as field stars solely based on their spatial distribution.
Due to external accretion, some stars kinematically associated with clusters might mistakenly be classified as field stars. 
Particularly for clusters like NGC 1783 with large cores, the limited field of view of the {\sl HST} can sometimes give the superficial impression that member stars are uniformly distributed in space.
Although the spatial distribution of this younger stellar population in NGC 1783 is less centrally concentrated than the bulk population, it still exhibits noticeable differences from that of field stars \citep{Li_stellar_2016}.
Their cumulative radial distribution is significantly more convex than a quadratic function representing field stars in Figure \ref{f3}(b).

The current study leads to the following conclusions:
\begin{enumerate}[label=(\arabic*)]
 \item The detected younger stellar population in \citetalias[]{li_formation_2016} are genuine cluster members. These younger stars are less centrally concentrated than the bulk population RGB and RC stars.
 \item The possibility that all these younger stars are BSSs formed through binary interactions can be ruled out.
 \item NGC 1783 might have undergone accretion involving the external, low-mass stellar system (including star clusters, stellar streams, and stellar associations), resulting in a mixture of external younger stars and BSSs from the older bulk population in the CMD. This confirms a minor merger scenario as proposed by \citet{hong_dynamical_2017}. 
\end{enumerate}

Considering the very active star formation in the MCs, we can further speculate that a considerable number of star clusters may have undergone similar accretion, some of which may still keep signatures of such processes. This could partly explain why \citet{2024Mohandasan} did not find any significant correlation between binary stars and BSSs in these clusters. 
In fact, both LMC and SMC host numerous binary (and multiple) star clusters being physically associated \citep{10.1093/mnras/230.2.215,1990A&A...230...11H}.
Many massive clusters, like NGC 1783, have experienced accretion and eventually became more massive than birth times.
For NGC 1783, the age difference between the younger population and the main body is too large to be originated from native matters, contradicting the classical theoretical formation model of gravitationally bound binary clusters through the splitting of the parent cloud.
If star clusters form in groups (Star Cluster Groups) inside giant molecular clouds, the likelihood of tidal capture forming binary clusters is significantly high due to the relatively high density of clusters there \citep[see, e.g.][and references therein]{leon1999interacting, 2007MNRAS.379..151M}.
Consequently, mergers are likely to be prevalent phenomenons, with many clusters competitively engulfing surrounding smaller stellar systems at their inception \citep[e.g.][]{2012ApJ...753...85F, 2017MNRAS.467.1313V, 2017MNRAS.472.4982S, 2018NatAs...2..725H, 2018MNRAS.481..688G, 2021MNRAS.502.6157C, 2022MNRAS.509..954D, 2023MNRAS.521.5557K}. 
However, these simulations only address the mergers of smaller clusters or subclusters forming nearly simultaneously within isolated molecular clouds. More simulations are needed to investigate the frequency of tidal capture forming massive clusters between systems with very different ages and masses \citep[e.g. NGC 1835 in][]{2023ApJ...953..125G} in large molecular complexes. 

Further investigation into newly formed clusters and their environments can significantly enhance our understanding of how star clusters form and evolve in starburst galaxies.

\begin{acknowledgements}
We thank the anonymous referee for the valuable comments and suggestions for improving our manuscript.
L. W. and C. L. are supported by the National
Key R\&D Program of China (2020YFC2201400). This work was supported by
the National Natural Science Foundation of China (NSFC) through grants
12073090 and 12233013. X.Y.P acknowledges financial support from the
NSFC through grant 12173029 and the research development fund of Xi'an
Jiaotong--Liverpool University (RDF-18--02--32). L.W. is grateful for
support from both the NSFC through grant 21BAA00619 and the
one-hundred-talent project of Sun Yat-sen University, the Fundamental
Research Funds for the Central Universities, Sun Yat-sen University
(22hytd09). R. d. G. acknowledges support from the Australian Research
Council Centre of Excellence for All Sky Astrophysics in 3 Dimensions
(ASTRO 3D), project number CE170100013.
\end{acknowledgements}

\vspace{5mm}

\software{{\tt\string Astropy} \citep{2013A&A...558A..33A,2018AJ....156..123A}, 
{\tt\string StarGO} \citep{Yuan_2018}, 
{\tt\string TOPCAT} \citep{taylor_topcat_2005},
{\tt\string PETAR} \citep{10.1093/mnras/staa1915}, 
{{\tt\string MCLUSTER} \citep{10.1111/j.1365-2966.2011.19412.x,10.1093/mnras/sty2232}}}

\bibliography{NGC1783}{}
\bibliographystyle{aasjournal}

\end{CJK*}
\end{document}